\documentclass[10pt]{iopart}
\usepackage{iopams}
\usepackage{graphicx}
\begin{document}

\title{Interferometry of direct photons in $^{208}$Pb+$^{208}$Pb collisions at 158 AGeV}

\author{D. Peressounko (for WA98 collaboration\footnote[1]{See \cite{WA98-dir} for the appropriate WA98 collaboration author
list.})}
\address{ RRC "Kurchatov Intitute", Kurchatov sq., 1, Moscow, 12312, Russia}
\ead{Dmitri.Peressounko@cern.ch}

\begin{abstract}
We present final results from the WA98 experiment which provide
first measurements of Bose-Einstein correlations of direct photons
in ultrarelativistic heavy ion collisions. Invariant
interferometric radii were extracted in the range $100<K_T<300$
MeV/c and compared to interferometric radii of charged pions. The
yield of direct photons for $100<p_T<300$ MeV/c was extracted from
the correlation strength parameter and compared to the yield of
direct photons measured in WA98 at higher $p_T$ with the
statistical subtraction method, and to predictions of a fireball
model.
\end{abstract}


Hanbury Brown-Twiss (HBT) interferometry provides a powerful tool
to explore the space-time dimensions of the emitting source
created in elementary particle or heavy ion collisions.
Historically, such measurements have concentrated on pion pair
correlations, but have also been applied to kaons, protons, and
even heavy fragments \cite{QM-rew}. Hadron correlations reflect
the space-time extent of the emitting source at the time of
freeze-out. Photon interferometry has several important
differences with respect to hadron interferometry. First, photons
having extremely large mean free path length provide the
possibility to measure directly the size of the innermost hottest
zone of the collision \cite{Srivastava}. Second, photons emitted
at different stages of the collision dominate  different regions
of transverse momentum, so, extracting correlation parameters in
different regions of average transverse momentum of the photon
pair, one can measure the space-time dimensions of different
stages of the collision. Here we present first measurements of
photon HBT correlations in ultrarelativistic heavy ion collisions.

A detailed description of the layout of the CERN experiment WA98
can be found in \cite{WA98-dir}. Here we briefly discuss those
subsystems used in the present analysis. The WA98 photon
spectrometer, comprising the LEad-glass photon Detector Array
(LEDA), was located at a distance of 21.5~m downstream from the
$^{208}$Pb target and provided partial azimuthal coverage over the
rapidity interval $2.35 < y < 2.95$. Further downstream, the total
transverse energy was measured in the MIRAC calorimeter. The total
transverse energy measured in MIRAC was used for offline
centrality selection. The analysis presented here was performed on
the 10\% most central $^{208}$Pb+$^{208}$Pb collisions at 158 AGeV
with a total sample of $5.8 \times 10^6$ events collected during
runs in 1995 and 1996. Similar analysis was performed on the 20\%
most peripheral collision data sample of  $3.9 \times 10^6$ events
but the statistical errors were at least an order of magnitude
larger than the expected signal.

\begin{figure}
\begin{center}
\includegraphics[width=6cm]{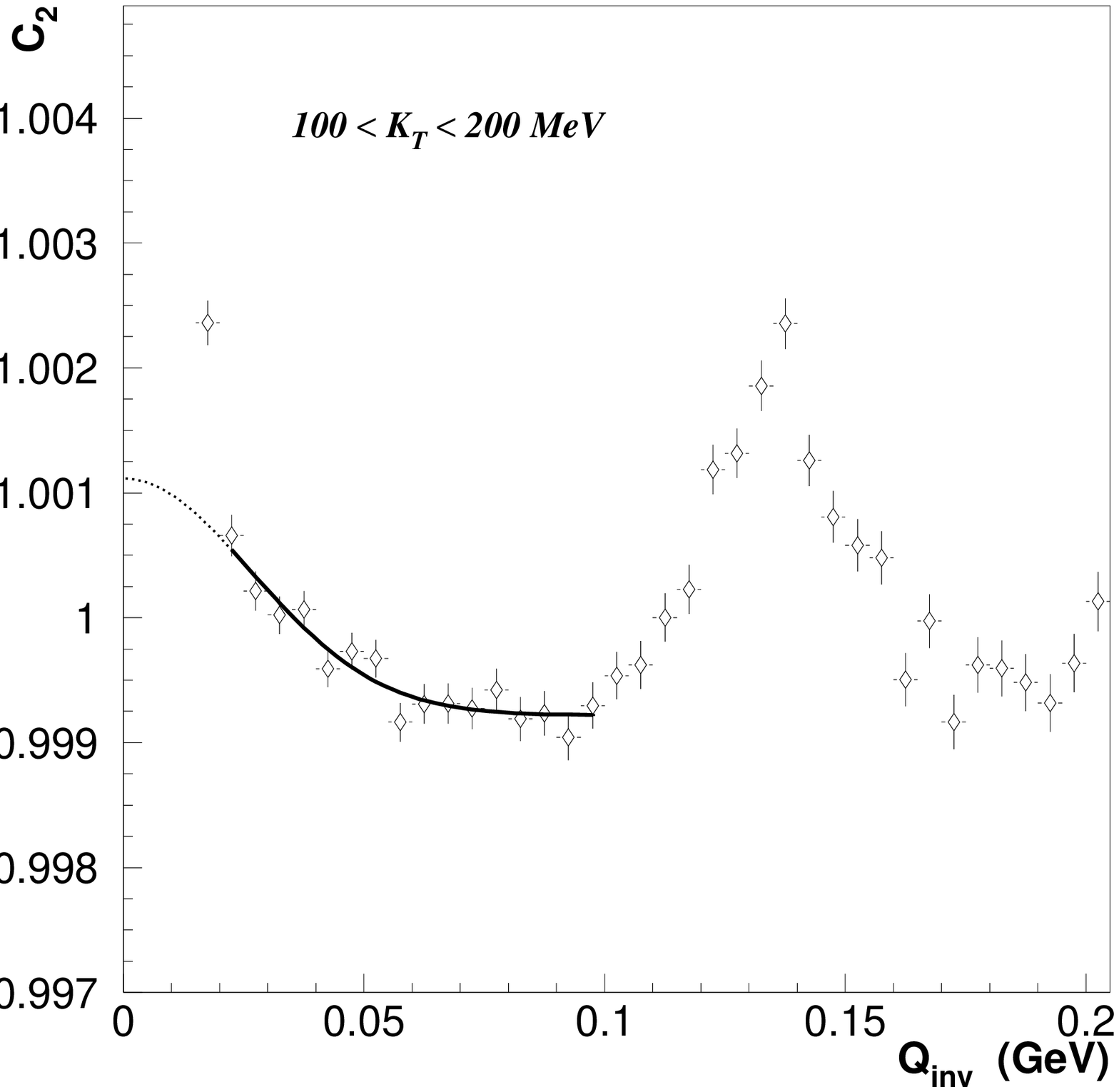}\quad
\includegraphics[width=6cm]{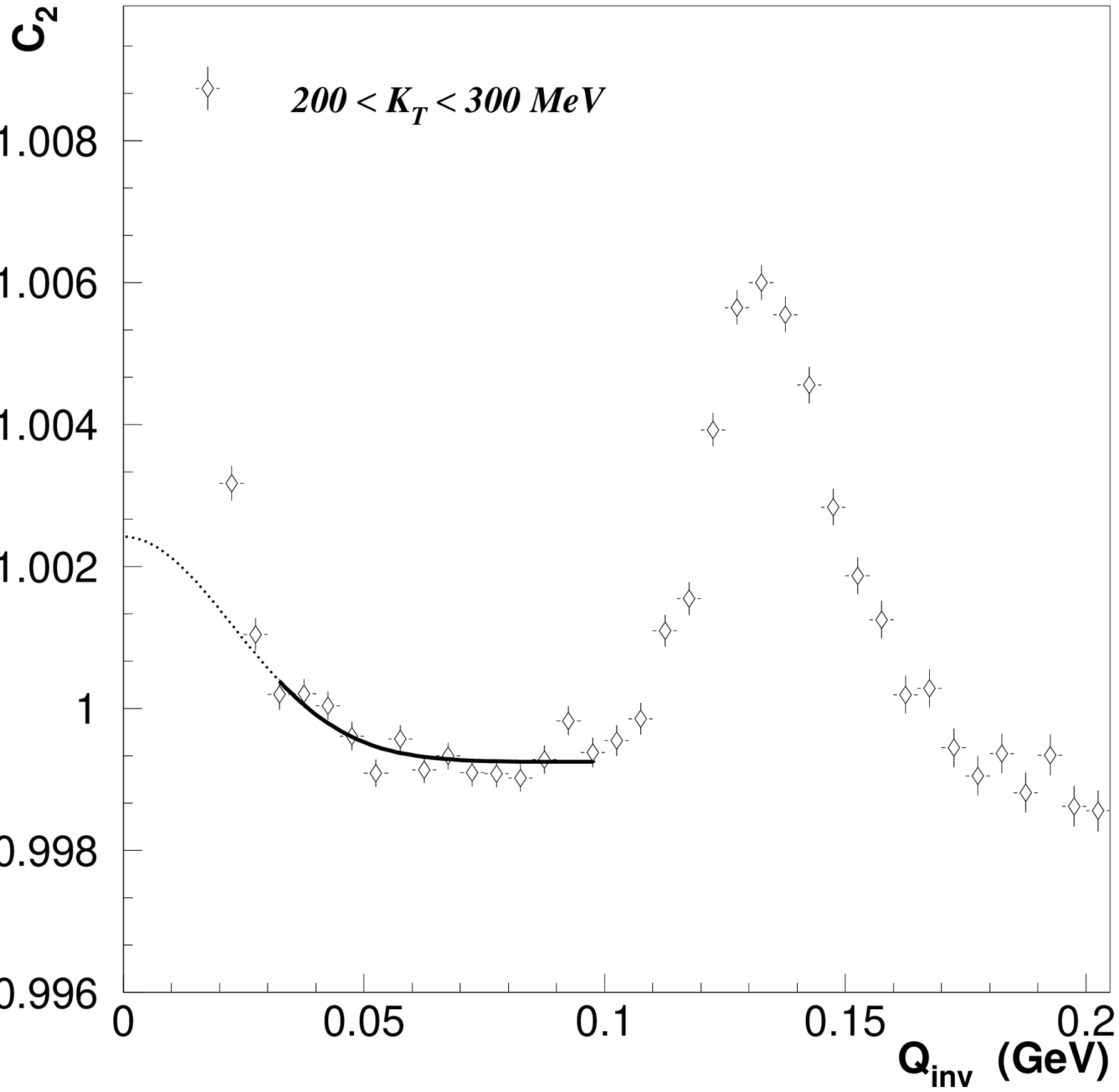}
\end{center}
\caption{\label{c2}Two-photon correlation functions calculated for
the 10 \% most central $^{208}$Pb+$^{208}$Pb collisions at 158
AGeV. "All" PID criterion is used and cut on minimal distance
$L_{12}>20$ cm is imposed.  }
\end{figure}

Examples of the two-photon correlation functions, extracted for
the 10 \% most central events are presented in  figure~\ref{c2}.
The small yield of direct photons and enormous background from
decays of final hadrons, mainly $\pi^0$, result in strength for
the two-photon correlations on the level of a tenth of a percent.
Measurement of such cor\-re\-la\-tions requires a good
understanding of the detector response and of possible
back\-gro\-unds. Possible sources of distortion of two-photon
correlation function are: 1) In\-ter\-fe\-ren\-ce of nearby
clusters in LEDA, erroneous splitting or merging of some clusters
by the reconstruction program; 2) Hadron misidentification; 3)
Photon conversion in front of LEDA; 4) Photon background
correlations, i.e. remnants of correlations of parent hadrons: HBT
correlations of parent $\pi^0$, elliptic flow, decays of
resonances.

To estimate the contribution of apparatus effects we constructed a
set of two-photon correlation functions, calculated with different
cuts on the minimal distance between clusters in LEDA. Then, we
fitted each varying the lower boundary of the fitting range and
compare the extracted correlation parameters. We find that for
distances between clusters larger than some minimal distance
$L_{12}^{min}\approx 20$~cm, depending on purity of the photon
spectrum, and/or for relative momenta larger than some minimal
relative momentum, also depending on the purity of the photon
spectrum, there is no dependence of the fit result on the minimal
distance cut or on the minimal relative momentum. That is,
correlations in this region are free from apparatus effects.

\begin{figure}
\begin{center}
\includegraphics[width=6.cm]{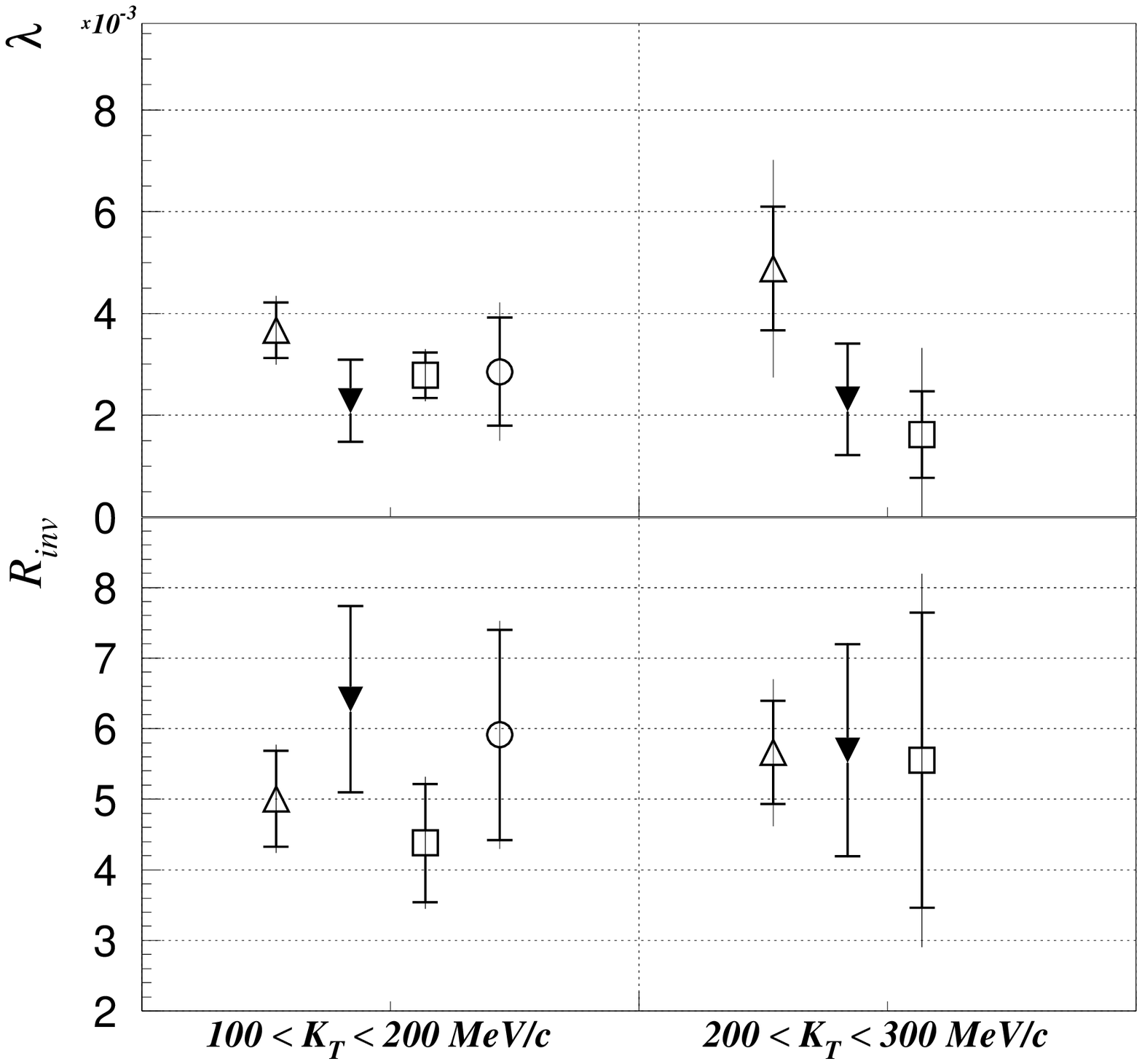} \quad
\includegraphics[width=6.cm]{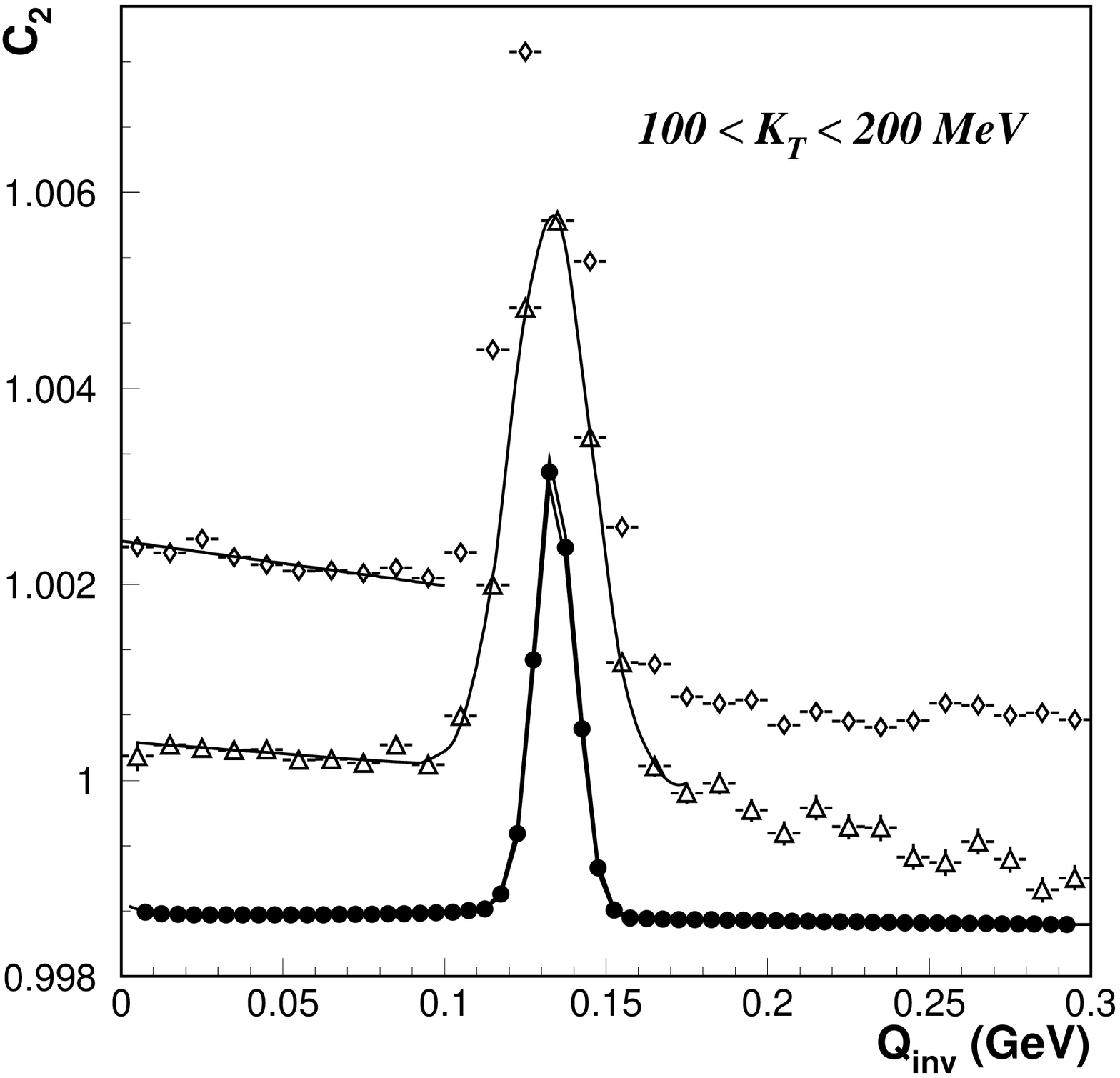}
\end{center}
\caption{\label{aparatus}Left plot: Comparison of the fit
parameters, corrected for efficiency and contamination, obtained
for different identification criteria:  $\triangle$ - "all",
$\blacktriangledown$ - "narrow", $\square$ - "neutral", $\circ$ -
"narrow neutral" (no significant result for high $K_T$). Right
plot: Simulated background photon correlations: $\diamond$ -
$\pi^0$ HBT correlations, $\triangle$ elliptic flow, $\bullet$ -
kinematic correlations (c.f. shifted vertically for clarity).}
\end{figure}

Contamination of the photon spectrum by charged particles and
neutral hadrons also can distort the photon correlation function.
Although hadrons deposit only a fraction of their energy, one
might still observe remnants of, e.g. their Bose-Einstein
correlations. Similarly, electron-positron pairs, from conversion
of photons between target and LEDA could also produce enhancement
at small relative momenta. To estimate contribution of these
effects we compared correlation parameters, evaluated from
correlation functions obtained with different identification
criteria and corrected for contamination of photon spectrum. The
contamination of charged particles varied with identification
criteria from 1 \% in "neutral" and "narrow neutral" to 16 \% in
"narrow" and 37 \% in "all" conditions. Similarly, the
contamination of neutral hadrons into the photon spectrum
decreases from 5\% ("all", "neutral") to 1 \% ("narrow", "neutral
narrow") when a cut is imposed on the width of the shower. Despite
the strong variation of the contamination, the corrected
correlation strengths, calculated for different identification
criteria coincide, see figure~\ref{aparatus}.

Background correlations, i.e. correlations between products of
decays of correlated hadrons, could result in enhancement at small
relative momenta. The dominant part of decay photons comes from
$\pi^0$ decays. We considered the following correlations:
Bose-Einstein $\pi^0$ correlations, elliptic flow, and
correlations due to decays of heavier resonances. To estimate
value of all these correlations, we performed Monte-Carlo
simulations. In all cases we used realistic rapidity and $p_t$
distributions, we take into account acceptance and energy and
position resolutions of the LEDA. We used Bose-Einstein
correlation parameters of $\pi^\pm$ as well as flow parameters of
charged pions. In the case of Bose-Einstein correlations we obtain
step-like correlations with small slope at small relative momenta,
in accordance with analytical calculations. Similarly, elliptic
flow results in the appearance of the even smaller slope at small
relative momenta. As for kinematic correlations, they are
completely negligible in this analysis. We find that all observed
correlations result in appearance of small slope at small relative
momenta. Calculating final values of parameters of two-photon
correlations we account for this slope.

\begin{figure}
\begin{center}
\includegraphics[width=6.cm]{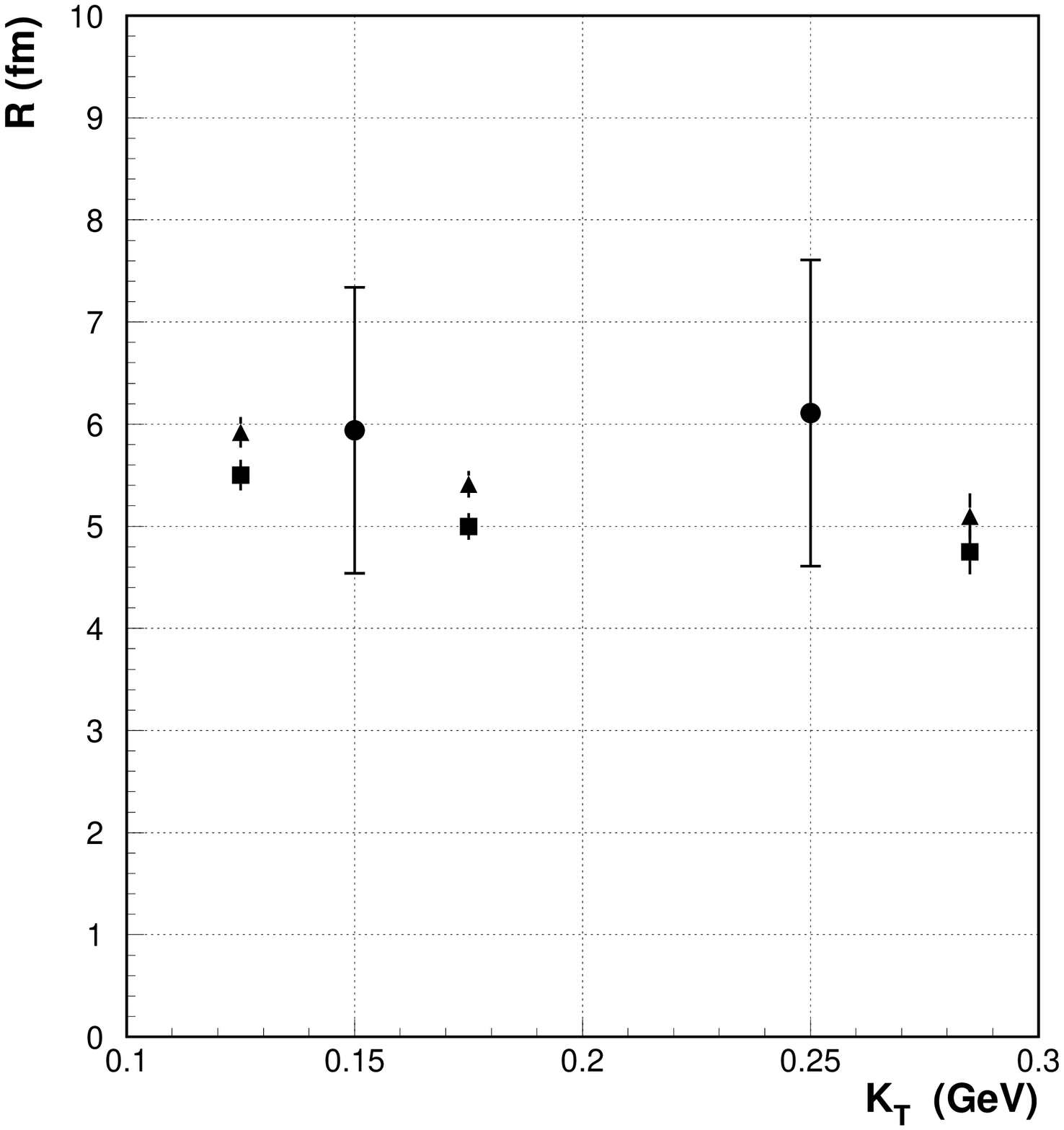} \quad
\includegraphics[width=6.cm]{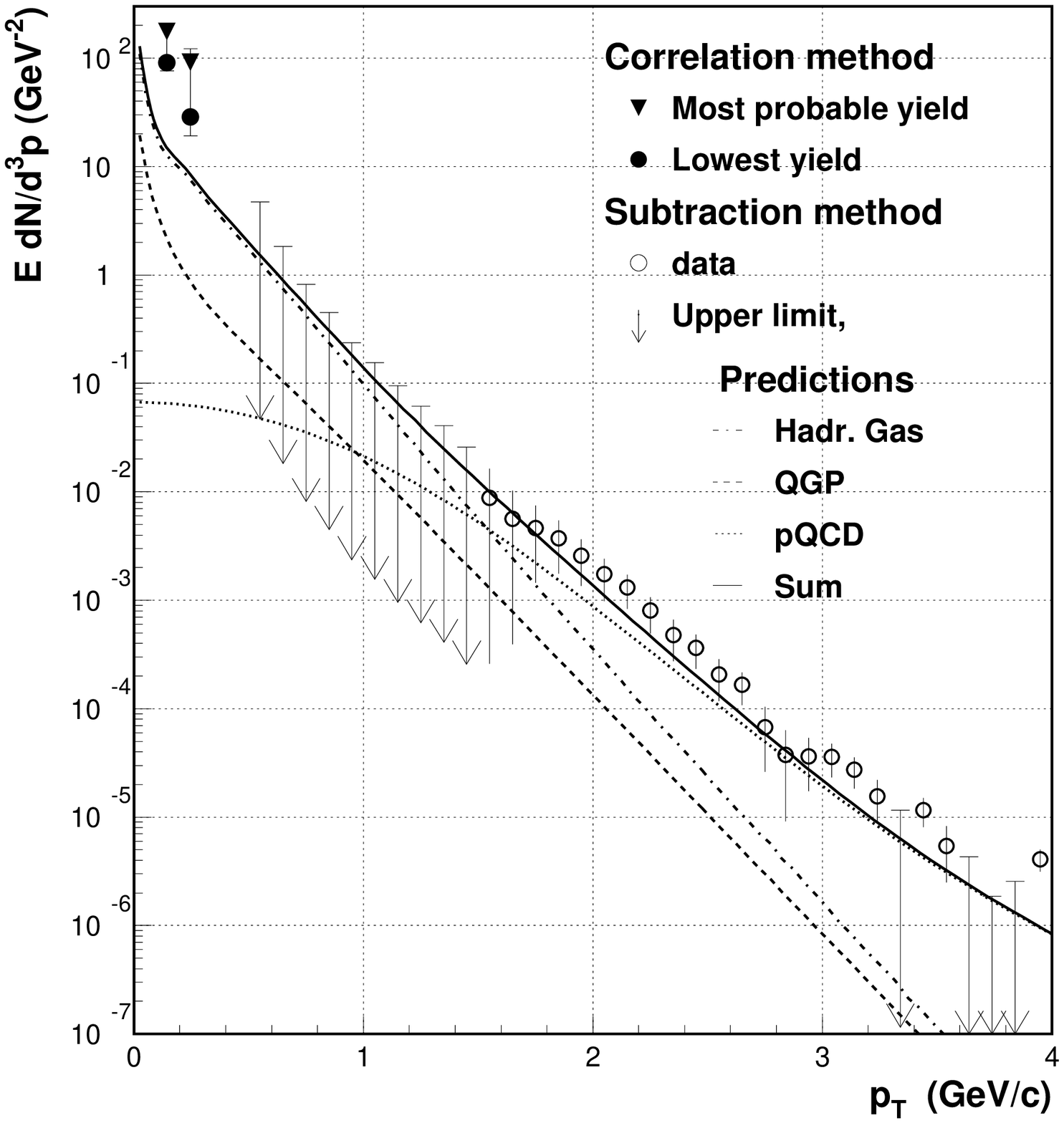}
\end{center}
\caption{\label{Result}Left plot: Comparson of photon invariant
correlation radius  ($\bullet$) with charged pion correlation
radii $R_{long}$ ($\blacktriangle$) and $R_{side}$
($\blacksquare$)  \cite{WA98-hm}. Right plot: Direct photon yield,
obtained with two methods and theoretical predictions \cite{Rapp}.
}
\end{figure}

As far as photons are massless particles, their invariant
correlation parameters have considerably different meaning than
those of hadron ones. One can express the correlation function on
invariant relative momentum as an integral over directions of
relative momenta in the center-of-mass frame of the pair:
\begin{eqnarray}
\label{params}
C_2(Q_{inv}) =1 + \frac{\lambda}{4\pi}\int do \exp \left \{- (Q_{inv}^2 + 4K_T^2)R_o^2\cos^2 \theta  \right . \nonumber \\
\left . - Q_{inv}^2 (R_s^2 \sin^2\theta \sin^2\phi + R_l^2 \sin^2\theta \cos^2\phi )  \right \},
\end{eqnarray}
Note that in this formula both correlation radii and projections
of relative momenta are taken in the local co-mover system. We
find that for massless particles the invariant correlation radius
is an average over {\it long} and {\it side} correlation radii and
that the invariant correlation strength decreases with respect to
the true correlation strength with increasing average transverse
momentum. Therefore, we should compare the photon invariant radius
to hadron {\it long} and {\it side} correlation radii, see
figure~\ref{Result}. We find that the direct photon correlation
radius is consistent with pion correlation radius.

Having the invariant correlation strength parameter, we extract
the direct photon yield. As we have seen in (\ref{params}), the
invariant correlation strength decreases with respect to the true
correlation strength with increase of $R_{out}\cdot K_T$. We have
not extracted the $R_{out}$ correlation radius, so we can only
find a lower limit on the direct photon yield, by assuming
$R_{out}=0$ and in addition - {\it a probable} direct photon
yield, corresponding to $R_{out}=6$ fm. We compare our result with
the result obtained earlier with the subtraction method and with
theoretical predictions, see figure \ref{Result}. We find that
results obtained with the correlation method are considerably
above theoretical predictions.

To conclude, the two-photon correlation function was measured for
the first time in ultrarelativistic heavy ion collisions. The
invariant correlation radius was measured at $100<K_T<300$~MeV/c.
The photon correlation radius is very similar to those of charged
pions. From the correlation strength parameter the lower limit on
the direct photon yield was extracted. Even lower limit appears to
be considerably larger than theoretical predictions.

\end{document}